\documentclass[11pt]{article}

\usepackage[final]{acl}
\usepackage{amsmath} 
\usepackage{graphicx}
\usepackage{subcaption} % 如果用子图的话

\usepackage{times}
\usepackage{latexsym}

\usepackage{array}     % for \arraybackslash and column modifiers
\usepackage{booktabs}  % for \toprule, \midrule, \bottomrule
\usepackage[T1]{fontenc}
\usepackage{todonotes}
\usepackage[utf8]{inputenc}

\usepackage{microtype}

\usepackage{inconsolata}

\usepackage{graphicx}
\usepackage{enumitem}
\usepackage{tabularx,booktabs}
\usepackage{siunitx}
\usepackage{xcolor}
\usepackage{colortbl}
\usepackage{amssymb}
\title{AutoDataset: A Lightweight System for Continuous Dataset Discovery and Search}

\author{
  \textbf{Junzhe Yang}\textsuperscript{1,2},
  \textbf{Xinghao Chen}\textsuperscript{1,3}, 
  \textbf{Yunuo Liu}\textsuperscript{1}, \textbf{Zhijing Sun}\textsuperscript{1}, \textbf{Wenjin Guo}\textsuperscript{1}, \textbf{Xiaoyu Shen}\textsuperscript{1} 
  \\
  \\
  \textsuperscript{1} Institute of Digital Twin, Eastern Institute of Technology, Ningbo \\
  \textsuperscript{2} Shanghai Jiao Tong University \quad  \textsuperscript{3} The Hong Kong Polytechnic University \\
  \texttt{yang\_j@sjtu.edu.cn}, \texttt{xyshen@eitech.edu.cn}
}

\begin{document}
\maketitle

\begin{abstract}
The continuous expansion of task-specific datasets has become a major driver of progress in machine learning. However, discovering newly released datasets remains difficult, as existing platforms largely depend on manual curation or community submissions, leading to limited coverage and substantial delays. To address this challenge, we introduce \textsc{AutoDataset}, a lightweight, automated system for real-time dataset discovery and retrieval. 
\textsc{AutoDataset} adopts a paper-first approach by continuously monitoring arXiv to detect and index datasets directly from newly published research. The system operates through a low-overhead multi-stage pipeline. First, a lightweight classifier rapidly filters titles and abstracts to identify papers releasing datasets, achieving an F1 score of 0.94 with an inference latency of 11\,ms. For identified papers, we parse PDFs with GROBID and apply a sentence-level extractor to extract dataset descriptions. Dataset URLs are extracted from the paper text with an automated fallback to \LaTeX{} source analysis when needed. Finally, the structured records are indexed using a dense semantic retriever, enabling low-latency natural language search.
We deploy \textsc{AutoDataset} as a live system that continuously ingests new papers and provides up-to-date dataset discovery. In practice, it has been shown to significantly reduce the time required for researchers to locate newly released datasets, improving dataset discovery efficiency by up to 80\%.\footnote{\textbf{System Availability:} The source code is publicly released at \url{https://github.com/EIT-NLP/AutoDataset}. A  screencast video demonstrating the complete workflow is available at \url{https://youtu.be/_QSxHKyIYns}.}
\end{abstract}

\section{Introduction}
Machine learning has undergone rapid advances in recent years, largely driven by the continued scaling of training data and model size~\citep{kaplan2020scaling,zhao2023survey,chung2024scaling,su2023welmwellreadpretrainedlanguage,su2024unraveling,deepseekai2025deepseekv32pushingfrontieropen}. In many subfields, progress is now tightly coupled to the availability of high-quality, task-specific datasets~\cite{haddow2022survey,shen2022lowresourcedenseretrievalopendomain,zhu2023meta,zhu2023weaker}. This dependency has fueled an explosion of dataset creation efforts, reflected in the steady rise of papers  releasing new datasets and benchmarks~\citep{lhoest2021datasets,liu2024llmdatasets}. 
As the pace of dataset publication accelerates, discovery has become increasingly difficult. While \emph{dataset search} has emerged as a critical research area in its own right~\citep{paton2024datasetdiscovery,li2025opendatasetsearch}, the sheer volume and velocity of new publications, as illustrated in Table~\ref{tab:dataset-growth}, now overwhelm manual curation and ad hoc tracking. These trends call for automated systems capable of \emph{tracking, structuring, and indexing} newly released datasets immediately upon publication, rather than waiting for manual curation.

\begin{table}[t]
\centering
\resizebox{\columnwidth}{!}{%
\begin{tabular}{lcccccc}
\toprule
\multicolumn{7}{c}{\textbf{Early Phase (2004--2014): Gradual Emergence}} \\
\midrule
\textbf{Year} & 2004 & 2006 & 2008 & 2010 & 2012 & 2014 \\
\textbf{Papers} & $\sim$5 & $\sim$10 & $\sim$5 & $\sim$15 & $\sim$40 & \cellcolor{gray!20}$\sim$90 \\
\midrule\midrule
\multicolumn{7}{c}{\textbf{Recent Phase (2020--2025): Explosive Growth}} \\
\midrule
\textbf{Year} & 2020 & 2021 & 2022 & 2023 & 2024 & 2025 \\
\textbf{Papers} & \cellcolor{purple!10}2,865 & \cellcolor{purple!10}2,479 & \cellcolor{purple!25}4,746 & \cellcolor{purple!40}7,347 & \cellcolor{purple!60}\textcolor{white}{7,655} & \cellcolor{purple!80}\textcolor{white}{\textbf{8,566}} \\
\textbf{YoY Growth} & -- & $-13\%$ & \textbf{+91\%} & \textbf{+55\%} & +4\% & +12\% \\
\textbf{vs. 2014} & 32$\times$ & 27$\times$ & 53$\times$ & 82$\times$ & 85$\times$ & \textbf{95$\times$} \\
\bottomrule
\end{tabular}%
}
% \caption{The rapid, explosive growth of dataset-introducing papers. By partitioning the timeline, the staggering shift from the early pre-2015 baseline to the modern era is clearly visible, culminating in an approximate 95-fold multiplier by 2025.Estimated number of dataset-introducing papers per year, as detected by our pipeline on arXiv}
\caption{\small Yearly counts of papers that our BERT-Gate detector classifies as introducing a new dataset, based on arXiv submissions in six computer science categories: IR, DB, AI, CL, CV and MA. The counts increase sharply from the pre-2015 baseline to recent years, with an $\sim$95-fold increase by 2025.}

\label{tab:dataset-growth}
\end{table}

To address this challenge, we introduce \textsc{AutoDataset}, a \emph{paper-first} automated system for \textbf{low-latency discovery and retrieval} of machine learning datasets. While community-driven hosting platforms such as Hugging Face~\cite{lhoest2021datasets} or OpenML~\cite{vanschoren2013openml} provide valuable infrastructure, their reliance on voluntary uploads often leads to inconsistent documentation, uneven maintenance, and severe deduplication issues, with the same dataset frequently appearing under different names, versions, or formats. In contrast, we build on a simple but powerful observation: high-quality machine learning datasets are typically introduced through peer-reviewed or preprint research papers. These papers provide rich, standardized signals, covering data collection, task formulation, experimental validation, and intended use that strongly indicate dataset quality and relevance. Moreover, because a dataset is usually associated with a single canonical introductory paper, anchoring discovery to publications naturally avoids redundancy without resorting to brittle heuristic-based deduplication.

Operationally, \textsc{AutoDataset} continuously monitors arXiv~\footnote{\url{https://arxiv.org/}}, the premier preprint hub for modern machine learning research, to ingest dataset-introducing papers. Unlike one-off document understanding tasks or offline curation, our system must process every newly submitted paper under strict latency and throughput constraints. This requirement makes naive reliance on large language model (LLM) agents impractical due to their high computational cost and operational overhead~\cite{guo2024large,shen2025mind,belcak2025smalllanguagemodelsfuture,ding2026llms}. Instead, \textsc{AutoDataset} adopts a deliberately \emph{lightweight, modular pipeline} to enable scalable ingestion suitable for long-term deployment. Specifically, the system utilizes a two-stage BERT-based architecture~\cite{devlin2019bert}. The first stage, \textsc{BERT-Gate}, functions as a high-speed filter that \emph{classifies} papers based solely on their titles and abstracts. We adopted this design because titles and abstracts reliably contain sufficient semantic signals to identify dataset contributions, allowing the system to accurately filter thousands of papers while maintaining a minimal processing time of roughly 11ms per paper.

For positively identified papers, the second stage, \textsc{BERT-Desc}, parses the full PDFs to \emph{extract} concise, human-readable dataset descriptions. Concurrently, to ensure the successful extraction of the primary dataset URL, the system extracts links directly from the PDF and automatically consults the \LaTeX~source when necessary. Finally, the structured records are \emph{indexed} using a state-of-the-art dense semantic retriever~\cite{karpukhin2020dpr,li2023gte}, exposing them through an intuitive natural language search interface. We deploy the system as a configurable web application that supports real-time crawling, parameter tuning, and immediate dataset exploration. By directly surfacing verified dataset links alongside concise descriptions, \textsc{AutoDataset} reduces the time required to locate newly released datasets by an estimated 80\%, transforming a previously manual and error-prone workflow into a streamlined retrieval experience.

In summary, our work makes three key contributions: (1) We introduce \textsc{AutoDataset}, a fully automated, paper-first system that continuously monitors arXiv to surface and index dataset-introducing papers. (2) We design a highly efficient two-stage BERT pipeline for rapid filtering and description extraction, combined with a URL module that ensures successful link extraction directly from PDFs, consulting \LaTeX~sources when necessary. (3) We build a central, searchable index and release a lightweight, interactive web demo to facilitate efficient, real-time dataset retrieval.
\section{Related Work}

\begin{table*}[t] % Use table* if you want it to span two columns, or just table for one column
\centering
\small
\resizebox{\textwidth}{!}{% Use \columnwidth if in a single column environment
\begin{tabular}{l l c c c}
\toprule
\textbf{Platform} & \textbf{Curation Method} & \textbf{Literature-Grounded} & \textbf{Update Latency} & \textbf{Metadata Consistency} \\
\midrule
Google Dataset Search & Automated Web Crawl & Mixed & Moderate & Heterogeneous \\
Hugging Face Datasets & Community Upload & Partial & Moderate to High & Moderate \\
Kaggle & Manual / Community & Rarely & High & Low \\
OpenML & Manual Upload & Partial & High & High \\
Papers with Code & Community Curation & Yes & Low to Moderate & Moderate \\
\midrule
\rowcolor{gray!10} \textbf{\textsc{AutoDataset} (Ours)} & \textbf{Automated Paper-Mining} & \textbf{Yes} & \textbf{Extremely Low} & \textbf{High} \\
\bottomrule
\end{tabular}%
}
\caption{Comparison of dataset discovery platforms. Unlike existing repositories that rely on manual curation or general web crawling, \textsc{AutoDataset} directly and automatically mines the literature for near-zero latency.}

\label{tab:dataset-platforms}
\end{table*}

\paragraph{Dataset Discovery Platforms}
Existing platforms for dataset discovery, such as Google Dataset Search \citep{noy2019datasetsearch} , Hugging Face Datasets~\citep{lhoest2021datasets}, Kaggle~\citep{kaggle}, OpenML\citep{vanschoren2013openml}, and Papers with Code~\citep{paperswithcode}, largely rely on manual uploads and community curation. As summarized in Table~\ref{tab:dataset-platforms}, these platforms face several common limitations, including inconsistent metadata, mixed dataset quality, and a significant lag between a paper's release and the dataset's availability on the platform. Beyond catalog-style repositories, recent work such as DataScout~\citep{lin2025datascout} explores interactive, AI-assisted dataset discovery interfaces, though these are still built on top of existing, manually curated repositories rather than directly mining the literature.

\paragraph{Automated Dataset Mention Extraction}
Earlier work in automated dataset discovery treated mention extraction primarily as a sequence-labeling problem. This includes two-stage pipelines that classify sentences before identifying mentions~\citep{kumar2021dataquest}, or BiLSTM-CRF models applied over full text~\citep{zeng2024datasetmention}. To support robust evaluation in this domain, DMDD~\citep{pan2023dmdd} introduced a large-scale benchmark corpus for dataset-mention detection. More recently, large language models (LLMs) and synthetic data have become the main workhorses for identifying and normalizing dataset references at scale~\citep{santos2025datagatherer,li2024ragingdme,solatorio2025monitoring}. Closest to our setting is ChatPD~\citep{xu2025chatpd}, which utilizes an LLM-driven pipeline to construct a paper-dataset network from the literature. In contrast, \textsc{AutoDataset} prioritizes ingestion speed and efficiency, replacing heavy LLM extraction with a lightweight, highly targeted BERT pipeline to enable continuous, real-time monitoring of preprint servers.

\begin{figure*}[!htbp] 
    \centering
    \includegraphics[width=\textwidth]{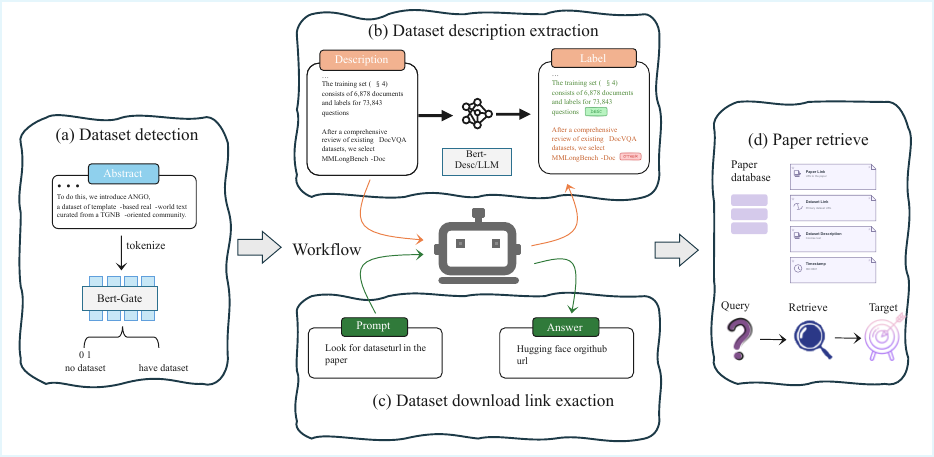}  
    \caption[System Architecture of AutoDataset]{System Architecture of AutoDataset}
    \label{fig:system-arch}
\end{figure*}

\section{The \textsc{AutoDataset} Pipeline}

To enable the continuous, low-latency ingestion of newly released datasets, \textsc{AutoDataset} is engineered as an efficient, end-to-end processing pipeline. As illustrated in Figure~\ref{fig:system-arch}, the system comprises four core modules: (1) the \textbf{Dataset Detection} module, which acts as a high-speed filter to identify dataset-introducing papers based on their titles and abstracts; (2) the \textbf{Description Extraction} module, which processes the full text of these positively identified papers to generate concise dataset summaries; (3) the \textbf{Link Extraction} module, which operates concurrently to reliably isolate the primary dataset URLs; and (4) the \textbf{Indexing and Retrieval} module, which consolidates the extracted metadata to power an intuitive natural language search interface. 

\paragraph{Dataset Detection}

To continuously monitor the literature, \textsc{AutoDataset} automatically ingests new arXiv submissions across six fundamental computer science domains known for high rates of dataset publication: Information Retrieval (IR), Database (DB), Artificial Intelligence (AI), Computational Linguistics (CL), Computer Vision (CV), and Multi-Agent Systems (MA). During this initial ingestion phase, paper metadata is robustly parsed using BeautifulSoup \cite{beautifulsoup} to handle complex HTML and XML document structures. Because the daily volume of new preprints is substantial, relying on heavy generative models for initial screening is computationally prohibitive. 

To solve this, we employ \textsc{BERT-Gate}, a lightweight binary classifier based on \texttt{bert-base-uncased}, to act as a high-speed initial filter. \textsc{BERT-Gate} processes only the concatenated title and abstract of each paper, which we found provides sufficient semantic signal to reliably determine if a new dataset is being introduced. 
The model was trained on a curated dataset of 7,789 annotated papers \footnote{Initial weak labels by Qwen-Plus were manually verified.}. Evaluated on a held-out test set, \textsc{BERT-Gate} demonstrates strong separability, achieving an accuracy of 0.97 and an F1 score of 0.94. Crucially, this lightweight architecture enables an average  processing latency of approximately 11ms per paper. Papers that exceed a prediction threshold of 0.5 are classified as positive and immediately passed to the Description Extraction module.\footnote{\label{fn:config} Appendix~\ref{sec:appendix-training-details} includes detailed training hyperparameters.}

\begin{table*}[t]
\centering

% 不再整体 \small，用正文默认字号更符合 ACL 建议
% --- 1. 纵向微调：略微压缩行距，但不要太狠 ---
\renewcommand{\arraystretch}{0.98}

% --- 2. 横向舒展：保持适度列间距，让表格在双栏中显得丰满 ---
\setlength{\tabcolsep}{9.5pt}

\begin{tabular}{l ccccc}
\toprule
\textbf{Model} & \textbf{Acc.} & \textbf{Prec.} & \textbf{Rec.} & \textbf{F1} & \textbf{Time (ms)} \\
\midrule
Longformer              & 0.923 & 0.678 & 0.831 & 0.747 & 265.20 \\
BigBird                 & 0.919 & 0.679 & 0.774 & 0.723 & 257.98 \\
Qwen-Plus (API)         & 0.809 & 0.636 & \textbf{0.933} & 0.757 & 1039.75 \\
\midrule
\textbf{BERT-Desc} 
& \textbf{0.949} {\footnotesize \textcolor{green!60!black}{$\uparrow$2.8\%}} 
& \textbf{0.800} {\footnotesize \textcolor{green!60!black}{$\uparrow$17.8\%}} 
& 0.924           {\footnotesize \textcolor{red!80!black}{$\downarrow$1.0\%}} 
& \textbf{0.858}  {\footnotesize \textcolor{green!60!black}{$\uparrow$13.3\%}} 
& \textbf{9.39}  {\footnotesize \textcolor{green!60!black}{$\downarrow$ 96.5\%}} \\
\rowcolor{gray!15} \quad - w/o window & 0.828 & 0.544 & 0.654 & 0.594 & 9.46 \\ 
\bottomrule
\end{tabular}

\caption{Performance on the dataset-description extraction task. The proposed \textbf{BERT-Desc} significantly outperforms baselines in overall F1 and efficiency. Colored percentages indicate the relative gain ($\uparrow$) or reduction ($\downarrow$) compared to the best-performing baseline in each respective metric. The bottom row presents the ablation study.}
\label{tab:model-comparison}
\end{table*}

\paragraph{Description Extraction}

Once \textsc{BERT-Gate} identifies a paper as dataset-introducing, the system initiates a deep-content analysis to extract a concise, machine-readable dataset description. To ensure robust text extraction, the system downloads the canonical PDF and processes it using GROBID ~\cite{lopez2009grobid} to generate structured, sentence-level XML representations. 

To determine which sentences actually describe the dataset, we employ \textsc{BERT-Desc}, a fine-tuned sentence-level classifier\textsuperscript{\ref{fn:config}}. Because dataset descriptions often span multiple sentences, classifying sentences in isolation severely limits performance. To address this, we developed a \emph{class-conditional sliding window} approach. For each target sentence, the model is fed a symmetric window of surrounding text up to its maximum token limit. To provide rich discourse context while efficiently down-weighting redundant negative spans, the window slides with a smaller stride around positive labels and a larger stride around negatives. 

% As detailed in Table~\ref{tab:model-comparison}, this contextualized approach achieves high accuracy and F1 scores, significantly outperforming the baseline that lacks contextual windows. Furthermore, \textsc{BERT-Desc} operates with minimal inference latency. It drastically outperforms heavy long-context document transformers, such as Longformer ~\cite{beltagy2020longformer} and BigBird ~\cite{zaheer2020bigbird}, in both predictive accuracy and processing speed. Similarly, it completely avoids the severe inference bottlenecks associated with API-based large language models. 

As detailed in Table~\ref{tab:model-comparison}, this contextualized approach achieves high accuracy and F1 scores, significantly outperforming the baseline that lacks contextual windows. Furthermore, \textsc{BERT-Desc} operates with minimal inference latency. It drastically outperforms heavy long-context document transformers, such as Longformer~\cite{beltagy2020longformer} and BigBird~\cite{zaheer2020bigbird}, in both predictive accuracy and processing speed. Similarly, it completely avoids the severe inference bottlenecks associated with API-based large language models, such as Qwen-Plus~\cite{qwen2023qwen}.

Sentences classified as positive are aggregated in document order to form the final dataset description. If \textsc{BERT-Desc} finds zero positive sentences in a document, the paper is re-classified as a negative. This allows the module to act as a highly effective second-stage filter that reduces false positives from the initial detection phase. In production, this entire extraction pipeline—spanning PDF downloading, parsing, and inference—processes documents rapidly enough to support continuous real-time ingestion.

\begin{figure}[!t]
    \centering
    % 左图: BERT-Gate
    \begin{minipage}{0.48\columnwidth}
        \centering
        \includegraphics[width=\linewidth]{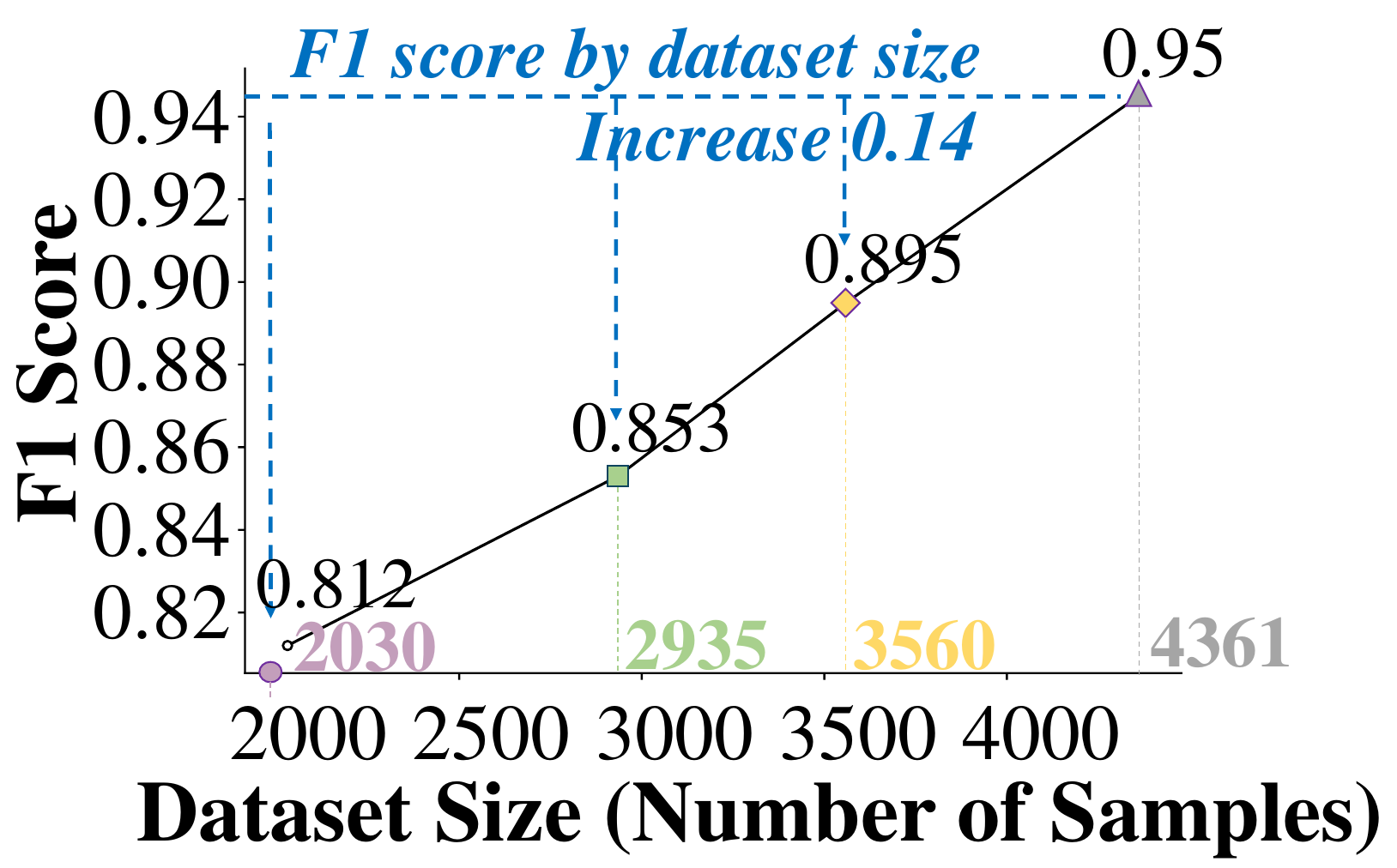}
        \caption*{(a) \textsc{BERT-Gate} F1 by dataset size.}
    \end{minipage}\hfill
    % 右图: BERT-Desc
    \begin{minipage}{0.48\columnwidth}
        \centering
        \includegraphics[width=\linewidth]{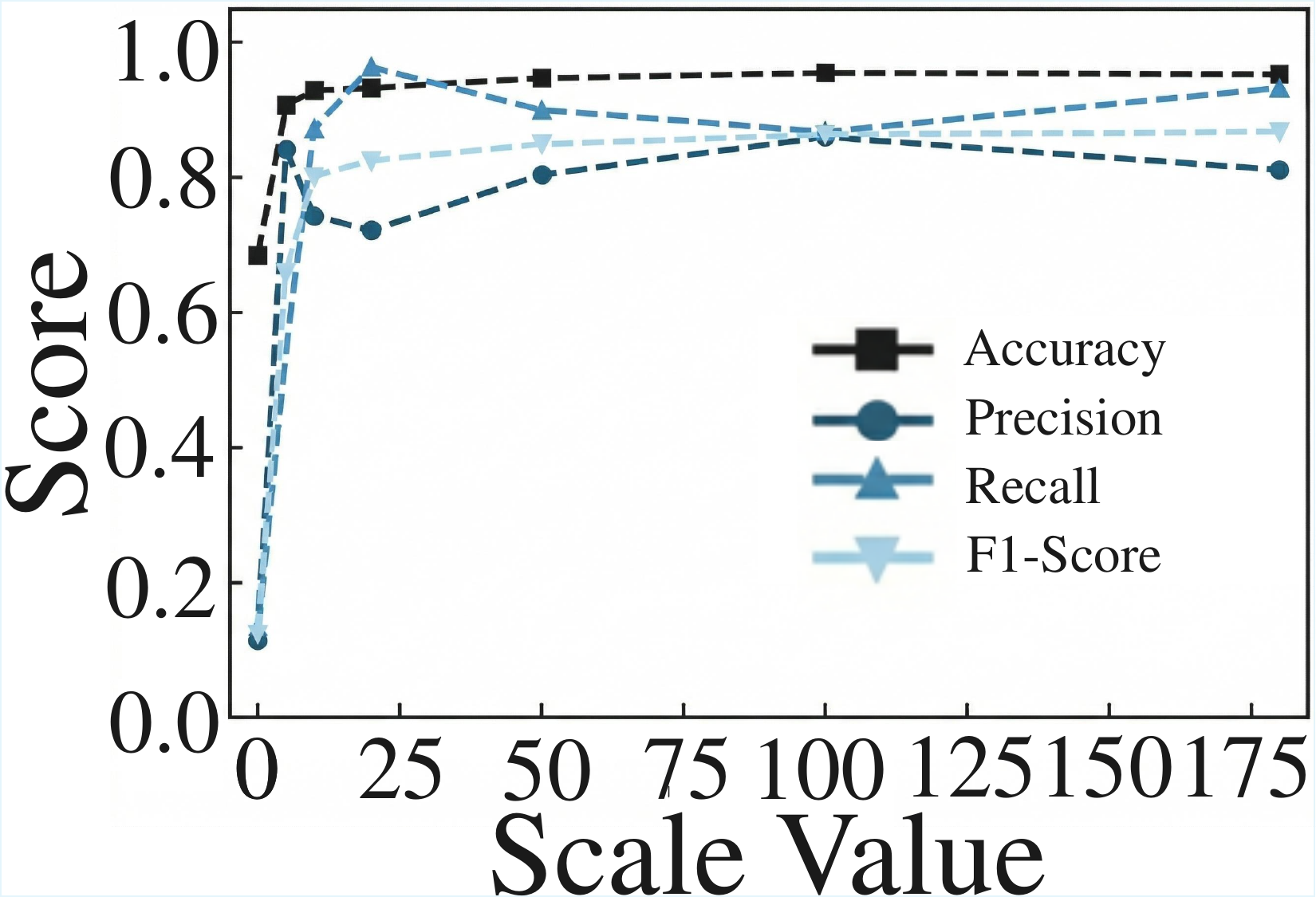}
        \caption*{(b) \textsc{BERT-Desc} metrics by dataset size.}
    \end{minipage}
    \caption{Data scaling ablation studies. Both models demonstrate strong data efficiency, rapidly approaching saturation (optimal performance) without requiring massive annotated corpora.}
    \label{fig:combined-scaling}
\end{figure}

%这个是并排放置的代码
\begin{figure*}[!t]
    \centering

    \begin{subfigure}{0.32\textwidth}
        \centering
        \includegraphics[width=\linewidth]{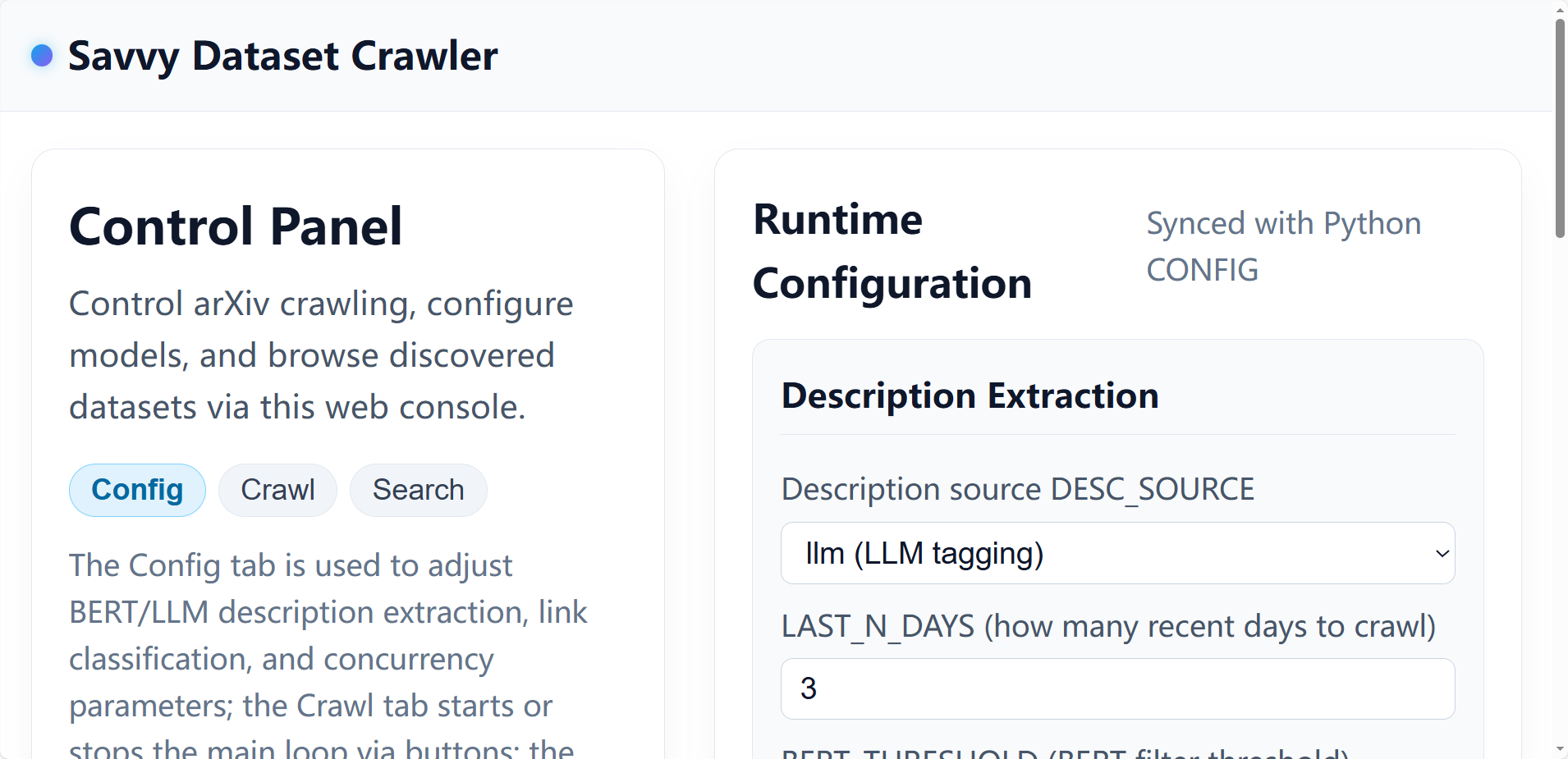}
        \caption{Configuration panel}
        \label{fig:demo-config}
    \end{subfigure}
    \hfill % 使用 hfill 自动填充水平间距
    \begin{subfigure}{0.32\textwidth}
        \centering
        \includegraphics[width=\linewidth]{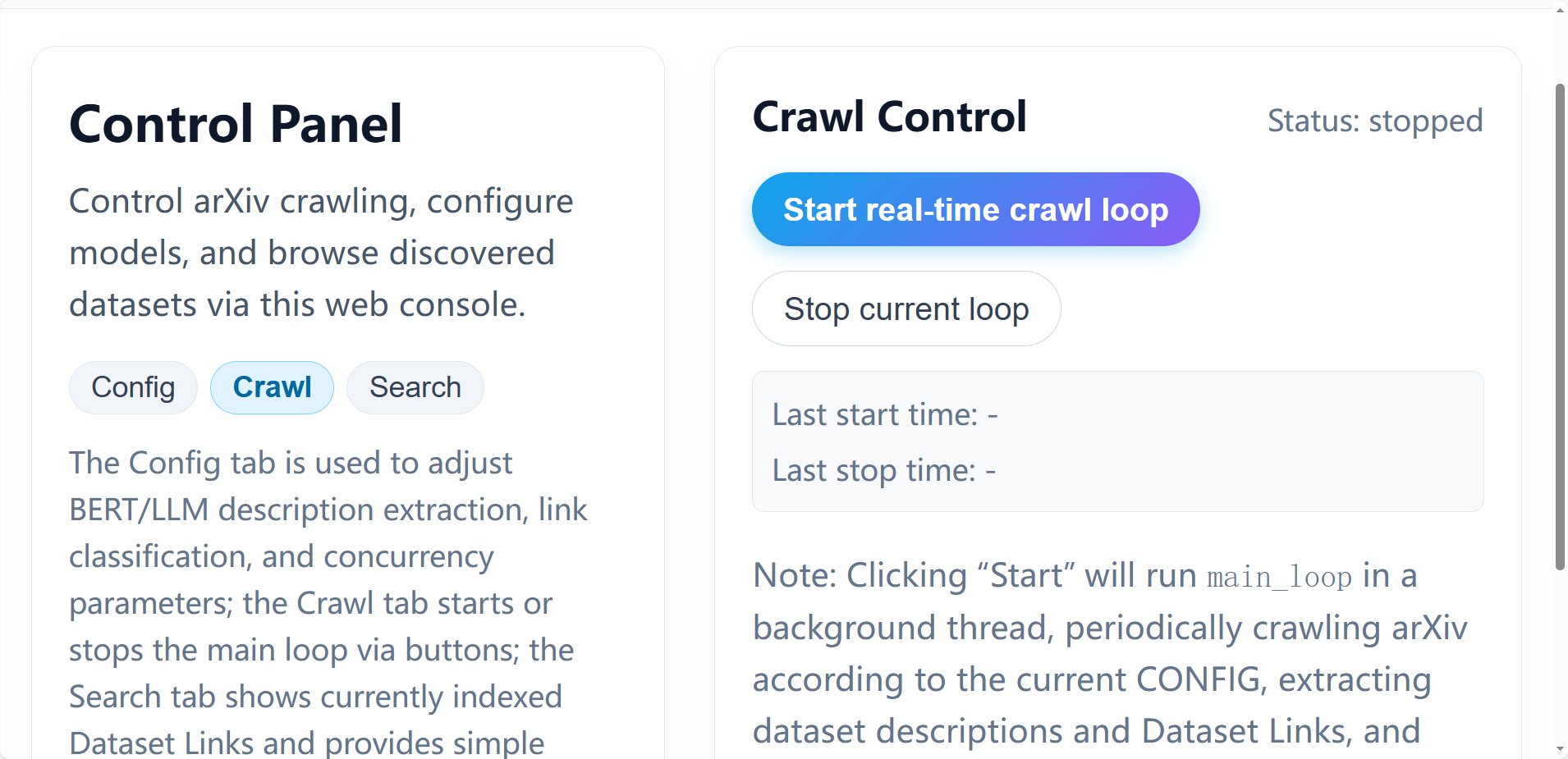}
        \caption{Crawl control panel}
        \label{fig:demo-crawl}
    \end{subfigure}
    \hfill
    \begin{subfigure}{0.32\textwidth}
        \centering
        \includegraphics[width=\linewidth]{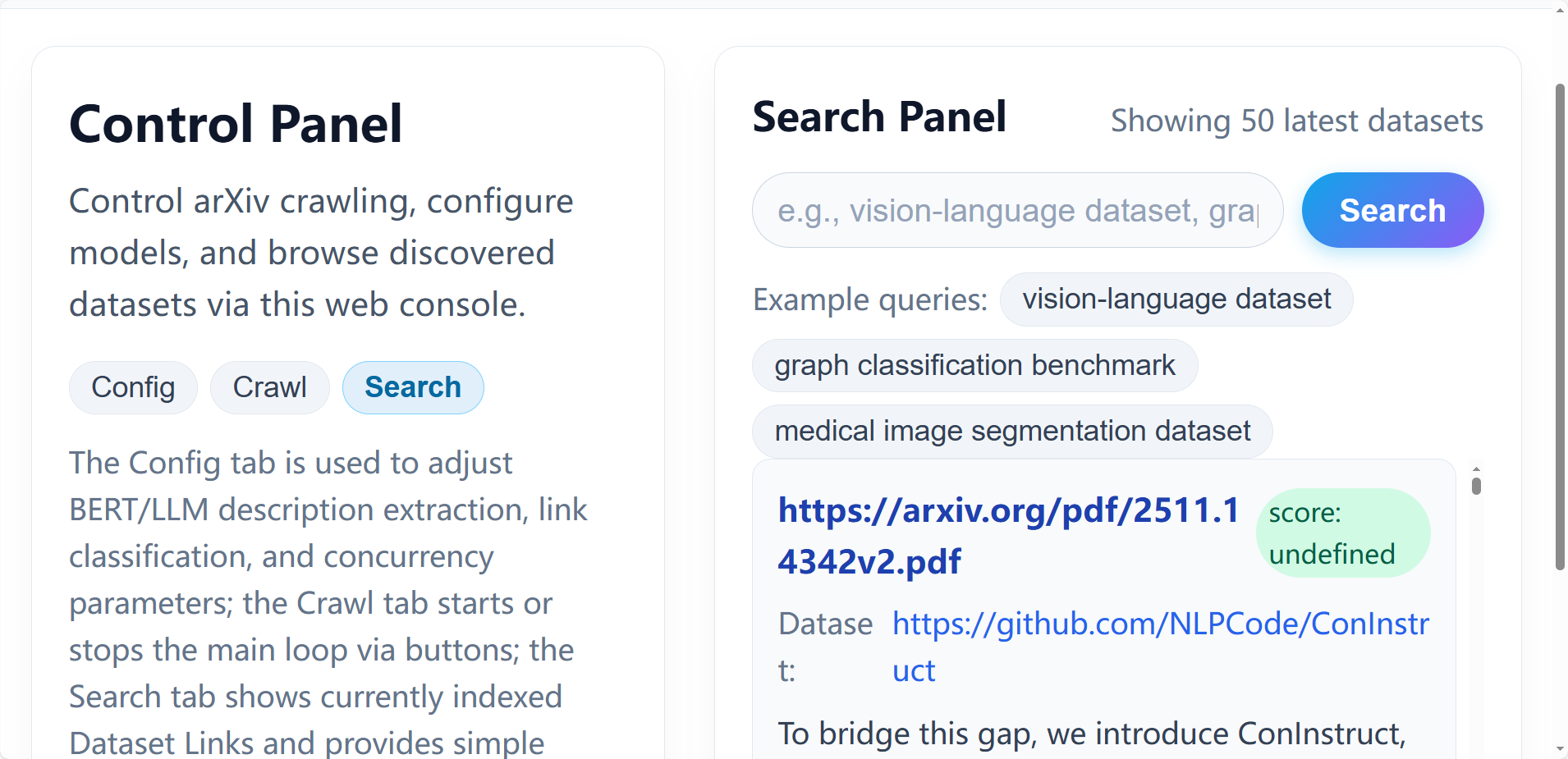}
        \caption{Search and retrieval panel}
        \label{fig:demo-search}
    \end{subfigure}

    \caption{The \textsc{AutoDataset} Web Interface. Users can dynamically configure extraction parameters (a), monitor the real-time arXiv ingestion loop (b), and explore the resulting dataset index via a dense semantic search (c).}
    \label{fig:demo-ui}
\end{figure*}

Both \textsc{BERT-Gate} and \textsc{BERT-Desc} exhibit exceptional data efficiency. As shown in Figure~\ref{fig:combined-scaling}, we conducted data scaling ablation studies by evaluating performance across varying training set sizes. Both models rapidly approach performance saturation with minimal annotated samples. This strong data efficiency validates our design choice: by targeting highly structured textual components, \textsc{AutoDataset} can be fully optimized without the prohibitive cost of massive human annotation.

\paragraph{Dataset Link Extraction}

Concurrent with description extraction, the system isolates the primary URL where the newly introduced dataset is hosted. Relying solely on parsed PDFs for link extraction often fails because URLs are frequently detached from their descriptive context, such as being relegated to footnotes or bibliographies. This dislocation severs the semantic connection necessary for accurate identification.

To guarantee successful link extraction, \textsc{AutoDataset} employs a robust, hybrid approach based on \LaTeX~source code analysis. Since the vast majority of papers in our target arXiv categories provide downloadable \LaTeX~sources, the system fetches these files and extracts all standard hyperlink occurrences. It then anchors each candidate URL with its surrounding sentences to preserve semantic context. 

% Following candidate extraction, we apply a comprehensive rule-based scoring system to filter out clearly irrelevant URLs, such as generic academic portals or root-level repository links. For the remaining candidates, an optional and lightweight language model verification step acts as a final classifier to confidently select the single primary dataset URL. If this verification module is disabled or returns uncertain results, the system seamlessly falls back to the top-scoring candidate from the rule-based heuristics. The final output is a structured JSON record containing the paper link, the primary dataset link, and the extracted description, all ready for downstream indexing.

Following candidate extraction, we apply a comprehensive rule-based scoring system (see Appendix~\ref{app:link-rules}) to filter out clearly irrelevant URLs, such as generic academic portals or root-level repository links. For the remaining candidates, an optional and lightweight language model verification step acts as a final classifier to confidently select the single primary dataset URL. If this verification module is disabled or returns uncertain results, the system seamlessly falls back to the top-scoring candidate from the rule-based heuristics. The final output is a structured JSON record containing the paper link, the primary dataset link, and the extracted description, all ready for downstream indexing.

\paragraph{Dataset Retrieval and Evaluation}

To bridge the extracted dataset records with user search intents, \textsc{AutoDataset} employs a dense semantic retrieval architecture~\citep{karpukhin2020dpr}. Both the user's natural language queries and the pipeline-extracted dataset descriptions are mapped into a shared high-dimensional vector space. For this embedding layer, we utilize the \texttt{Alibaba-NLP/gte-large-en-v1.5} encoder~\citep{li2023gte}, selected for its robust semantic search capabilities. To guarantee low latency at query time, all dataset descriptions are pre-encoded offline. When a user submits a search request, the query is embedded in real-time and matched against the offline index using cosine similarity to return the most relevant nearest neighbors.

To evaluate the system's end-to-end retrieval efficacy, we curated a benchmark of natural language queries, each mapping to a specific ground-truth dataset record. As illustrated in the search interface (Figure~\ref{fig:demo-search}), the retriever demonstrates highly accurate surfacing capabilities, consistently ranking the correct dataset at the very top of the search results. Expanding the retrieval window slightly further yields nearly perfect coverage. These metrics confirm that the system successfully bridges the semantic gap between informal user queries and the automatically extracted descriptions, delivering a highly responsive and precise search experience.

\section{Interactive Web Demonstration}
\label{sec:demo}

%A live demonstration of the \textsc{AutoDataset} system is available at \url{https://your-system-link.com}.

To demonstrate the practical utility of \textsc{AutoDataset}, we deployed the system as a lightweight, interactive Flask web application. As illustrated in Figure~\ref{fig:demo-ui}, the interface is organized into three distinct panels that allow users to manage the entire paper-to-dataset pipeline: a \textbf{Configuration} panel, a \textbf{Crawl} controller, and a natural language \textbf{Search} interface.

\paragraph{System Configuration and Crawling}
The configuration panel (Figure~\ref{fig:demo-config}) exposes the pipeline's core hyperparameters through an intuitive form. Users can dynamically adjust the BERT filtering confidence thresholds, specify the time window of arXiv papers to monitor, and toggle the use of the LLM fallback for link verification. System-level parameters, such as the number of parallel worker threads and maximum PDF download concurrency, can also be adjusted at runtime to accommodate different hardware constraints.
Once configured, users can monitor the automated ingestion process via the Crawl panel (Figure~\ref{fig:demo-crawl}). As the system polls arXiv and processes new papers through the two-stage BERT pipeline, all extracted dataset records, including the descriptions and verified URLs, are serialized as JSON objects and continuously appended to the backend index.

% \paragraph{Dataset Search and Time Efficiency}
% The Search panel (Figure~\ref{fig:demo-search}) provides the primary interface for exploring the discovered resources. Users enter natural language queries, which the backend encodes in real-time using the \texttt{gte-large-en-v1.5} model to retrieve the most semantically relevant datasets from the index. The results are rendered as a ranked list, displaying the source paper link, the extracted dataset description, and crucially, the verified primary dataset URL. 

% This direct surfacing of verified links offers significant workflow improvements. Traditionally, discovering a newly introduced dataset requires manually identifying candidate papers, scanning the PDF or appendix for mentions, and verifying the download link—a process that typically takes 4--9 minutes per query. By centralizing this process and immediately providing the verified dataset URL alongside its descriptive context, \textsc{AutoDataset} reduces this workflow to simple result inspection (roughly 0.5--2 minutes). This represents an estimated 60\% to 80\% reduction in the time-to-first-valid-link, proving especially valuable for accessing brand-new datasets that have not yet been cataloged by community platforms.

\begin{figure}[t]
    \centering
    \includegraphics[width=\linewidth]{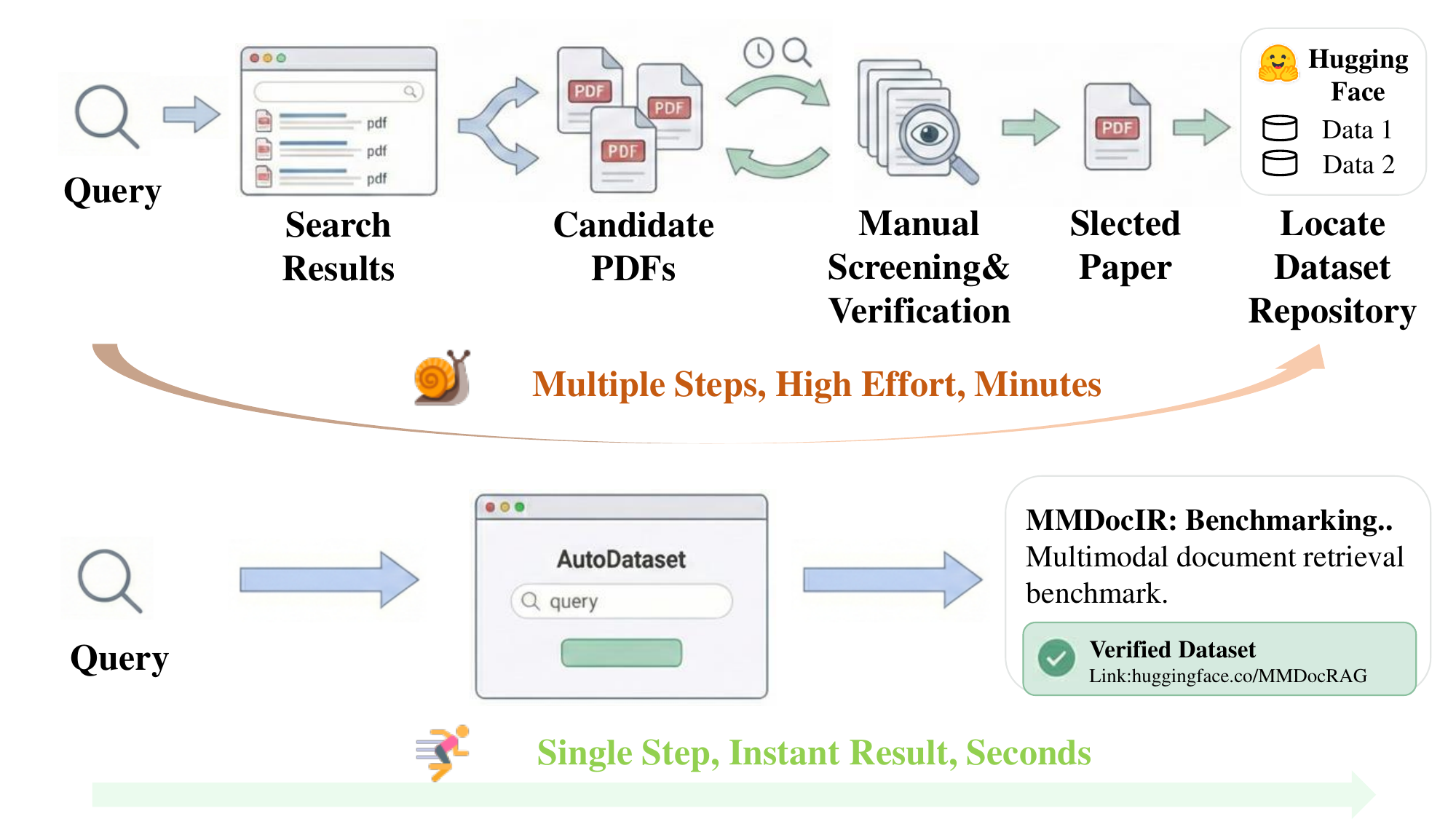} % 不写 .pdf 也可以
    \caption{Comparison between the traditional manual workflow and the \textsc{AutoDataset} workflow for discovering multimodal document information datasets.}
    \label{fig:workflow-comparison}
\end{figure}

\section{Case Study}
To illustrate how \textsc{AutoDataset} supports realistic research workflows, we consider a user who is interested in newly released datasets for \emph{multimodal document information retrieval}. A natural starting point is to issue a Web query such as ``multimodal document information'' in a general search engine and manually search for a suitable benchmark.

Figure~\ref{fig:workflow-comparison} contrasts this interaction with the traditional manual workflow.
Without \textsc{AutoDataset}, the user first issues a Web search, inspects the result snippets, opens several candidate arXiv PDFs, and then scans the main text or appendix for dataset mentions. For each candidate paper, they must follow project links (e.g., GitHub or Hugging Face), and manually verify which URL corresponds to the actual benchmark repository. In the MMDocIR case, this involves navigating from the search engine to the arXiv PDF of the paper ``Benchmarking Multimodal Retrieval for Long Documents'', then locating the MMDocIR project page and, finally, the dataset repository.

With \textsc{AutoDataset}, the same information need is handled much more directly.
The user submits the query ``multimodal document retrieval'' in the Search panel, and the system returns a dataset–centric record for the MMDocIR benchmark, linking to the source paper~\footnote{\url{https://arxiv.org/pdf/2501.08828v3.pdf}} and to the predicted primary dataset URL~\footnote{\url{https://mmdocrag.github.io/MMDocIR/}}, together with a short textual description and a timestamp indicating when the record was last observed in the crawl.
We manually verified that the returned URL is the official project page, which provides an overview of the benchmark, task definitions, and access to both the evaluation and training sets.

This direct surfacing of verified links offers significant workflow improvements.
Based on a small-scale pilot study with three researchers searching for five recently published datasets, discovering a newly introduced dataset manually usually requires 4--9 minutes per query, dominated by PDF scanning and link verification.
In contrast, using \textsc{AutoDataset} reduces the workflow to inspecting a single ranked result and clicking a verified dataset URL, which typically takes only 0.5--2 minutes.
This corresponds to an estimated 60--80\% reduction in time-to-first-valid-link, and is especially valuable for accessing very recent datasets that have not yet been catalogued by community platforms.

\section{Conclusion and Future Work}

In this paper, we introduced \textsc{AutoDataset}, a fully automated, paper-first system designed to continuously monitor the preprint literature and index newly released machine learning datasets with near-zero latency. By replacing heavy generative models with a highly efficient two-stage BERT pipeline and robust \LaTeX-assisted link extraction, the system achieves rapid processing speeds without sacrificing precision. The accompanying web demonstration proves that this automated, literature-grounded approach is both technically feasible and highly practical for accelerating dataset discovery.

In future work, we plan to enhance the system by extracting more fine-grained metadata---such as task applicability, data modality, and language tags—directly from the text. We also aim to implement stronger cross-paper deduplication and tighter integration with existing community repositories like Hugging Face and Papers with Code. Ultimately, \textsc{AutoDataset} establishes a robust foundation for building richer, real-time discovery services that keep pace with the rapid evolution of empirical machine learning.

\medskip

% \section*{Limitations}
% Our system has several practical constraints. First, monitoring only specific arXiv computer science categories omits datasets released elsewhere. Second, relying on GROBID-parsed text makes our pipeline sensitive to PDF parsing failures and domain shifts. Third, while our BERT-based gate is efficient, its current training primarily covers English-language publications; extending support to multilingual research remains future work. Finally, the demo employs a single-stage dense retriever over short descriptions without exploiting richer structured metadata. We view these as necessary design choices for a lightweight demo, aiming to provide immediate practical utility while ensuring high system responsiveness.

% \section*{Acknowledgments}
% %此处需要补充
% We thank EIT and IDT High Performance Com puting Center for providing computational resources for this project.

% % Bibliography entries for the entire Anthology, followed by custom entriesf

\bibliographystyle{acl_natbib}
\bibliography{custom}

\newpage
\appendix

\section{Model Training Details and Scaling Analysis}
\label{sec:appendix-training-details}

\paragraph{Data Preparation and Hyperparameters}
Both the \textsc{BERT-Gate} and \textsc{BERT-Desc} models were fine-tuned using the configurations detailed in Table~\ref{tab:hyperparameters}. All experiments were conducted on a single NVIDIA RTX 3090 (24\,GB) GPU. 

For \textsc{BERT-Gate}, the classifier was trained on 7,789 instances split into training, validation, and test sets using a 10:3:3 ratio, meticulously maintaining a 1:2.8 positive-to-negative sample ratio across all splits. For \textsc{BERT-Desc}, the model was trained on a meticulously annotated corpus of 179 papers comprising 21,297 sentences, and evaluated on a held-out test set of 2,286 sentences. Additionally, \textsc{BERT-Desc} utilized masked mean pooling followed by a two-layer MLP (Linear--LayerNorm--Dropout 0.3--GELU--Linear) to produce the final sentence-level logits, and its cross-entropy loss was configured with \texttt{ignore\_index} $=-100$.

% \begin{table}[h]
% \centering
% \label{tab:hyperparameters}
% \small
% \begin{tabular}{lcc}
% \toprule
% \textbf{Hyperparameter} & \textbf{\textsc{BERT-Gate}} & \textbf{\textsc{BERT-Desc}} \\
% \midrule
% Max Sequence Length     & 512 & 512 \\
% Optimizer               & AdamW & AdamW \\
% Learning Rate           & $3 \times 10^{-5}$ & $2 \times 10^{-5}$ \\
% Weight Decay            & 0.1 & 0.01 \\
% Batch Size              & 16 & 1 (Effective: 4) \\
% Evaluation Strategy     & Every 100 steps & Epoch-based \\
% Model Selection Metric  & Validation Loss & Validation Recall \\
% \bottomrule
% \label{tab:config}
% \end{tabular}
% \caption{Training configurations and hyperparameters for the two BERT-based modules.}
% \end{table}

\begin{table}[h]
\centering
\small
\begin{tabular}{@{}lcc@{}} % 注意这里的 @{}
\toprule
\textbf{Hyperparameter} & \textbf{\textsc{BERT-Gate}} & \textbf{\textsc{BERT-Desc}} \\
\midrule
Max Sequence Length     & 512 & 512 \\
Optimizer               & AdamW & AdamW \\
Learning Rate           & $3 \times 10^{-5}$ & $2 \times 10^{-5}$ \\
Weight Decay            & 0.1 & 0.01 \\
Batch Size              & 16 & 1 (Effective: 4) \\
Evaluation Strategy     & Every 100 steps & Epoch-based \\
Model Selection Metric  & Validation Loss & Validation Recall \\
\bottomrule
\end{tabular}
\caption{Training configurations and hyperparameters for the two BERT-based modules.}
\label{tab:hyperparameters}
\end{table}

\paragraph{Context Window Optimization (\textsc{BERT-Desc})}
To implement the class-conditional sliding window for the description extractor, we seeded a symmetric window with $\pm2$ sentences around the target sentence, expanding outwards until reaching the 512-token limit. The window slides with a smaller stride ($\lfloor \text{window}/3 \rfloor$, minimum 1) around positive labels and a larger stride ($\lfloor \text{window}/2 \rfloor$) elsewhere. This specific engineering choice enriches the local context for actual dataset descriptions while aggressively down-weighting redundant negative spans.

\paragraph{Data Scaling Analysis}
To assess the data efficiency of both architectures, we evaluated their performance across increasingly larger subsets of training data. 

As illustrated in Figure~\ref{fig:combined-scaling}(a), the F1 score for \textsc{BERT-Gate} rises monotonically with dataset size, progressing from 0.812 (at 2,030 samples) up to 0.945 (at 4,361 samples). This demonstrates that the detector is highly data-efficient even at lower resource levels and is highly sufficient for our use case as a lightweight filter. 

Similarly, as shown in Figure~\ref{fig:combined-scaling}(b), all four primary metrics for \textsc{BERT-Desc} improve rapidly when scaling from small to moderate dataset sizes. The curves then flatten significantly as they approach our full training scale, indicating that the sentence-level classifier is currently operating in a near-saturated regime where additional annotated data would yield negligible performance gains.

%第二部分
\section{Rule-Based Scoring for Dataset URLs}
\label{app:link-rules}

Given a candidate URL $u$, its anchor text $a$, and local context $c$
(two sentences before and after the hyperlink), we compute a scalar
score $\mathrm{score}(u)$ by summing a small set of interpretable
features. The final score is

\[
  \mathrm{score}(u) = \sum_i w_i f_i(u, a, c),
\]

where each $f_i \in \{0,1,2,\dots\}$ is a hand-crafted indicator or
count feature and $w_i$ is a small integer weight. Table~\ref{TAB:LINK-FEATURES}
summarizes the main feature groups and representative conditions.

\begin{table*}[t!]
\centering
\small
\begin{tabularx}{\textwidth}{l >{\raggedright\arraybackslash}X r}
\toprule
\textbf{GROUP} & \textbf{CONDITION} & \textbf{WEIGHT} \\
\midrule
HOST$^+$ & \texttt{HUGGINGFACE.CO/DATASETS} IN HOST+PATH & +10 \\
& \texttt{ZENODO.ORG/RECORD}, \texttt{KAGGLE.COM/DATASETS} & +8--9 \\
& \texttt{FIGSHARE.COM}, \texttt{DATAVERSE.ORG}, \texttt{OSF.IO} & +7--8 \\
\midrule
HOST$^-$ & \texttt{ARXIV.ORG}, \texttt{DOI.ORG}, \texttt{ACM.ORG}, & $-8$ TO $-10$ \\
& \texttt{IEEEXPLORE.IEEE.ORG}, \texttt{SCHOLAR.GOOGLE.*} & \\
& \texttt{RESEARCHGATE.NET}, \texttt{MEDIUM.COM} & $-6$ \\
\midrule
PATH HINTS & PATH CONTAINS \texttt{/DATASET}, \texttt{/DATASETS} & +3 \\
& PATH CONTAINS \texttt{/DATA}, \texttt{/DOWNLOAD}, & +2 \\
& \texttt{/FILES}, \texttt{/RECORD}, \texttt{/RELEASES} & \\
\midrule
FILE EXT & ENDS WITH \texttt{.CSV}, \texttt{.TSV}, \texttt{.JSON}, & +6 \\
& \texttt{PARQUET} & \\
& ENDS WITH \texttt{.ZIP}, \texttt{.TAR(.GZ)}, \texttt{.TGZ}, & +5 \\
& \texttt{.XZ}, \texttt{.7Z} & \\
& ENDS WITH \texttt{.RAR} & +4 \\
\midrule
LEXICAL$^+$ & IN $(A,C)$: ``DATASET'', ``OUR DATASET'', & +2 EACH \\
& ``WE RELEASE'', ``AVAILABLE AT'' & (CAP +8) \\
LEXICAL$^-$ & IN $(A,C)$: ``CODE'', ``SOURCE CODE'', & $-2$ EACH \\
& ``IMPLEMENTATION'', ``BIBTEX'' & (CAP $-6$) \\
SPECIAL$^-$ & IN $(A,C)$: ``WE EVALUATE ON'' AND ``DATASET'' CO-OCCUR & $-3$ \\
\midrule
GITHUB PENALTY & \texttt{GITHUB.COM} REPOSITORY ROOT WITHOUT & $-4$ \\
& \texttt{/RELEASES} OR \texttt{/DATA} IN PATH & \\
\bottomrule
\end{tabularx}
\caption{INTERPRETABLE INTEGER-WEIGHTED FEATURES FOR SCORING CANDIDATE DATASET URLS.}
\label{TAB:LINK-FEATURES}
\end{table*}

\subsection{Primary URL Selection Modes}
\label{app:link-selection}

To accommodate different computational constraints and accuracy requirements, our system supports three distinct operational modes for final link selection:

\begin{enumerate}[leftmargin=*, itemsep=4pt]
  \item \textbf{Rule-Only Mode.} This mode relies purely on the heuristic scoring system described above. Let candidates be ranked by decreasing score: $(u_1, s_1), (u_2, s_2), \dots$. We apply the following selection logic:
  \begin{itemize}
      \item \emph{High-confidence acceptance:} If there is only one candidate, we return it. Otherwise, we accept $u_1$ if $s_1 \ge \tau_{\text{high}}$ or $(s_1 \ge \tau_{\text{mid}} \;\wedge\; s_1 - s_2 \ge \Delta)$, where $\tau_{\text{high}}{=}22$, $\tau_{\text{mid}}{=}16$, and $\Delta{=}5$.
      \item \emph{Preferred-host tie-break:} If inconclusive, we examine the top-$K$ candidates ($K{=}5$) and prefer those hosted on dataset-first platforms (e.g., Hugging Face, Zenodo). We break ties by preferring landing pages over direct file links.
      \item \emph{General tie-break \& Rejection:} If no preferred-host URL is found, we select the highest-scoring candidate, using shorter URLs to break ties. If the best score $s_1 < \tau_{\text{min}}$ (where $\tau_{\text{min}}{=}15$), we mark the record as ``no reliable dataset link''.
  \end{itemize}
  While highly efficient, this purely rule-based scheme can occasionally misclassify complex cases.

  \item \textbf{LLM-Only Mode.} This mode bypasses the rule-based scoring entirely and delegates candidate selection to a Large Language Model (LLM). It prioritizes semantic understanding over extraction speed, relying on the LLM to comprehend the context around each link.

  \item \textbf{Hybrid Mode (Rule + LLM).} This is our default and most robust approach. It leverages the rule-based system as a first-stage filter to eliminate obvious non-dataset links (e.g., generic academic portals) based on their negative scores. The remaining high-quality candidates, along with their compact semantic contexts, are then passed to the LLM. The LLM acts as the final classifier to confidently pinpoint the primary dataset URL. If the LLM is disabled or returns uncertain results, the system seamlessly falls back to the top-scoring candidate from the Rule-Only logic.
\end{enumerate}

In practice, the integer weights and thresholds for the rule-based component were tuned via manual inspection on a development set of arXiv papers. This modular design ensures that the pipeline is easy to debug while maintaining high extraction precision.

\end{document}